\documentclass{article} 
\usepackage[a4paper, total={6.5in, 9in}]{geometry}
\usepackage{amsmath,amssymb}
\usepackage{natbib}
\usepackage{cite}
\usepackage{graphicx}
\usepackage{authblk} 
\usepackage{amsmath, bm}
\usepackage{enumitem}
\usepackage[ruled,vlined]{algorithm2e}
\usepackage{booktabs}
\usepackage{subfigure}

\title{Joint inversion for Vp, Vp/Vs of the San Fransico Bay Area using ADTomo}
\author[1,2,3,4]{Ling Xia}
\author[2,3]{Weiqiang Zhu}
\author[4]{Huajian Yao}
\affil[1]{Department of Geosciences, Princeton University}
\affil[2]{Department of Earth \& Planetary Science, University of California, Berkeley}
\affil[3]{Berkeley Seismology Laboratory, University of California Berkeley
}
\affil[4]{School of Earth and Space Sciences, University of Science and Technology of China
}
\date{\today}

\begin{document}
\maketitle

\begin{abstract}
This article presents a new seismological tomography method based on the fast sweeping method and advanced seismic phase picking techniques to study the complex geological structures of the San Francisco Bay Area. 

  By calculating the eikonal equation using the fast-sweeping method, this study obtains travel time information and gradient data under a given velocity structure. With an automatic differentiation algorithm to calculate gradients of the loss function and the L-BFGS algorithm to achieve optimization, the velocity model is iteratively adjusted to minimize the loss function. The P wave to S wave velocity ratio obtained through joint inversion is more reliable than the velocity ratio obtained directly by dividing the P wave and S wave velocity models. Compared to traditional inversion, the velocity ratio here does not require the same P wave and S wave travel path, thus improving the accuracy of the velocity ratio.

  The method was applied to the San Francisco Bay Area, a region with complex geological structures and significant seismic activities. This region, a focal point of research interest, also provides abundant seismic data for this study. We use deep learning methods to automatically pick seismic phases, acquiring abundant P wave and S wave travel time information. This project obtains the S wave velocity structure and P wave to S wave velocity ratio results for the first time. Compared to previous inversion results, the P wave velocity model here demonstrates higher resolution and better geological correspondence. High and low velocity anomalies also align well with geological maps, providing reasonable explanations in terms of lithology.
\end{abstract}

\section{Introduction}
Seismic tomography, using waveform data of natural earthquakes to inverse the underground three-dimensional structure, is a vital focus in geophysics. Accurate P and S wave velocity model shows clear underground structures, thereby revealing regional and global tectonic evolution(like \citet{hawley2016tomography,schmandt2010complex,stern2015seismic}), fault morphology(like \citet{allam2014seismic}, \citet{mcguire2005high}, \citet{ben1998properties}), and assisting earthquake relocation(like \citet{hauksson2012waveform}). Travel time tomography picks up the first arrival time of seismic waves, with earthquakes’ occurrence time, thus obtaining the propagation time of specific seismic phases. To inverse the velocity model, people use methods including high-frequency ray-path solutions(like \citet{gajewski1987computation}, \citet{cerveny2001seismic}), numerical solutions based on the eikonal equation(like \citet{biondi1992solving}, \citet{fomel2009fast}), and finite frequency imaging(like \citet{allam2014seismic}, \citet{hung2004imaging}). Traditionally, it is necessary to manually pick up the first arrival travel time data, which is time-consuming and requires more waveform data. With the advancement of machine learning(\citet{mousavi2023machine}, \citet{mousavi2022deep}), people have developed effective methods to automatically pick up the first arrival time(like \citet{zhu2019phasenet}, \citet{liu2020rapid}, \citet{park2020machine}, \citet{chai2020using}, \citet{tan2021machine}), thus providing sufficient datasets for tomography. This article utilizes PhaseNet(\citet{zhu2019phasenet}) to pick P wave and S wave phase data automatically, solves the eikonal equation based on the fast sweeping method, and finally obtains a detailed velocity model of the San Francisco Bay Area. 

Located along the boundary between the Pacific Plate and the North American Plate, the San Francisco Bay Area exhibits significant seismic activity, with complex fault systems like the San Andreas Fault and Hayward Fault. This place is also prone to large-magnitude earthquakes due to fault zones. Many people have done research here to investigate the subsurface velocity structure through different methods such as travel-time tomography (like \citet{thurber2007three}, \citet{hardebeck2007seismic}, \citet{hole2000three}, \citet{dorbath1996seismic}), finite-frequency tomography (like \citet{pollitz2007finite}), and surface wave tomography (like \citet{li2018rayleigh}). USGS also conducted sufficient research on this area(\citet{aagaard2020science}) and released a three-dimensional seismic velocity model from detailed geologic information(like \citet{aagaard2024usgs}). These velocity structure results help people a lot in earthquake modeling on large earthquakes here. 

From previous velocity structure results, people already have a better comprehension of P wave velocity model or S wave velocity models, while the inversion of the P wave to S wave velocity ratio is limited in this region. However, the P wave to S wave velocity ratio is especially vital in fault zone regions because it can reveal underground water properties. Individual travel time tomography can inverse both P wave and S wave velocity models using corresponding travel time data, reaching a high resolution. After PhaseNet(\citet{zhu2019phasenet}) provides sufficient S wave picks, S wave inversion with Vp/Vs ratio should be more applicable. With the development of machine learning, automatic differentiation has become a familiar technique to calculate the gradients. We hope to use these updated tools for tomography and reduce numerical artifacts as much as possible, thus developing a new method. 

This paper presents a novel travel time tomography method, automatic differentiation(AD) tomography, and its application in the San Francisco Bay Area, showcasing its unique contribution to the field. 
With the application of PhaseNet(\citet{zhu2019phasenet}), this research obtains datasets of abundant P wave and S wave phase picks easily, thus enabling the inversion of both P wave velocity, S wave velocity, and Vp/Vs ratio results. The P wave to S wave velocity ratio result was obtained through a joint inversion process, which utilized both P wave and S wave travel time to inverse both the P wave velocity model and the Vp/Vs ratio. The joint inversion result, which is more accurate than the direct division of P wave and S wave velocity models, does not assume the same travel path of P wave and S wave either. 
The improved quantity and quality of the dataset make it possible to reach a smaller inversion resolution; parallel programming with open MPI enables a larger inversion area. This method can also be applied to other places and effectively solve tomographic problems.

In this thesis, we first provide preliminary knowledge on travel time tomography. Following this, we describe the specific methods in tomography and relocation. Finally, we present the results of the P velocity model and the ratio of comprehensional to shear wave velocity model, with some explanations and analysis. 
\section{Preliminary Knowledge}

The tomography and relocation parts are mainly based on the travel time field, using the fast sweeping method to solve the eikonal equation. This chapter will go through the development of the eikonal equation, and the introduce the fast sweeping to calculate this equation. 

\subsection{Eikonal Equation}

From Continuous Media Mechanics, people use the equation of motion to describe ground movement.
\begin{equation}
    \rho \frac{\partial^2 u}{\partial t^2} = ( \lambda + 2\mu)\nabla(\nabla \cdot u) - \mu \nabla \times \nabla \times u
\end{equation}

If we use the divergence-free function and the curl-free function to describe the displacement function,
\begin{equation}
    u = \nabla \phi + \nabla \times \psi
\end{equation}

then equation(2.1) can be written into two parts after the departure of the divergence-free and curl-free parts.
\begin{equation}
    [\frac{\partial^2 }{\partial t^2} - \alpha ^2 \nabla^2] \phi = 0, \quad \alpha ^2 = (\lambda + 2\mu) / \rho
\end{equation}
\begin{equation}
    [\frac{\partial^2 }{\partial t^2} - \beta ^2 \nabla^2] \psi = 0,  \quad \beta ^2 = \mu / \rho
\end{equation}

We can summarize these two equations into one case.
\begin{equation}
    \nabla^2 \phi = \frac{1}{c(x,y,z)^2} \cdot \ddot{\phi}
\end{equation}

As this is a wave equation, the function of wave propagation should be.

\begin{equation}
\phi = \phi_0(x,y,z) \cdot \exp\left\{i \omega \left[\frac{r(x,y,z)}{c(x,y,z)}-t\right]\right\}
\end{equation}

The term $\omega \left[\frac{r(x,y,z)}{c(x,y,z)}-t\right]$ can describe the phase of this propagating wave. As the time increases, the related $\frac{r(x,y,z)}{c(x,y,z)}$ should also increase to remain the phase unchanged. Then, we can use another variable $\tau(x,y,z)$ to describe $\frac{r(x,y,z)}{c(x,y,z)}$, which could represent the constant phase wavefront. 
\begin{equation}
\phi = \phi_0(x,y,z) \cdot \exp\left\{i\omega \left[\tau(x,y,z)-t\right]\right\}
\end{equation}

Applying this function to the original equation, we can calculate the exact forms of the wave equation further. For simplification, we use $E$ to represent $\exp\left\{i\omega \left[\tau(x,y,z)-t\right]\right\}$. The left side can be calculated using the following process.
\begin{align}
     \frac{\partial \phi}{\partial x} &= (\frac{\partial \phi_0(x,y,z)}{\partial x}  + \phi_0(x,y,z) (i\omega)  \frac{\partial \tau(x,y,z)}{\partial x} ) \cdot  E
    \\
    \frac{\partial^2 \phi}{\partial^2 x} &= [\frac{\partial ^2 \phi_0}{\partial^2 x} -\phi_0 \omega^2(\frac{\partial \tau}{\partial x})^2+ 2i\omega \frac{\partial \tau}{\partial x}\frac{\partial \phi_0}{\partial x} + i\omega \phi_0 \frac{\partial^2 \tau}{\partial^2 x}] \cdot E
    \\
    \nabla ^2 \phi &= [\nabla ^2 \phi_0 -\phi_0 \omega^2 grad^2 \tau + i \omega (2\nabla \tau \cdot \nabla \phi_0 + \phi_0 \nabla ^2 \tau)]\cdot E
\end{align}

As to the right side of equation(2.6),
\begin{equation}
    \ddot \phi = -\omega ^2 \phi = -\omega ^2 \phi_0 \cdot E
\end{equation}

thus we can access the equation of $\tau(x,y,z)$. The real part and imaginary part of this equation should be the same independently. Thus, we can obtain these two equations.
\begin{align}
    -\frac{\omega ^2}{c(x,y,z)^2} \phi_0 &= \nabla^2 \phi_0 - \phi_0 \omega^2 grad^2 \tau
    \\[8pt]
    0 &= 2\nabla \tau \cdot \nabla \phi_0 + \phi_0 \nabla ^2 \tau
\end{align}

Therefore, in the high-frequency approximation case, the term $\nabla ^2 \phi_0$ can be neglected, and then we obtain the eikonal function, where $\tau$ represents the travel time of the wavefront.

\begin{equation}
    grad^2 \tau = \frac{1}{c(x,y,z)^2}
\end{equation}

\subsection{Fast Sweeping Method}

The eikonal equation provides a vital perspective to calculate the wavefront travel time. Based on the eikonal equation, this thesis uses the fast sweeping method to obtain the travel time field(refer to \citet{zhao2005fast}). To be more specific, we can rewrite the equation(2.14) to this form.
\begin{equation}
    \frac{\partial^2 \tau}{\partial^2 x} + \frac{\partial^2 \tau}{\partial^2 y} + \frac{\partial^2 \tau}{\partial^2 z}  = \frac{1}{c(x,y,z)^2}
\end{equation}

To solve it, we can discretize the partial differential equation at the integral grid point. For a more straightforward description, we use $u$ to represent travel time and $f$ to represent the slowness of propagating waves. Therefore, the discrete equation could be written as,
\begin{equation}
    [(u_{i,j,k}-u_{xmin})^+]^2 + [(u_{i,j,k}-u_{ymin})^+]^2 + [(u_{i,j,k}-u_{zmin})^+]^2 = f_{i,j,k}^2 h^2
\end{equation}

in which $\Delta u ^+$ represents the maximum of $\Delta u$ and 0; $u_{xmin}$ represents a minimum value between $u_{i-1,j,k}$ and $u_{i+1,j,k}$, so do $u_{ymin}$ and $u_{zmin}$. Based on this equation, we can derive the travel time in one specific grid. Sorting the travel time data in ambient grids, we use $a$, $b$, $c$ to represent $u_{xmin}$, $u_{ymin}$, $u_{zmin}$ with $a<b<c$.

\begin{enumerate}[label = \textbf{Case \arabic* :}, itemindent = 3em]

  \item $(u_{i,j,k}-a)^2 = f_{i,j,k}^2  h^2$ 
        \quad  $u_{i,j,k} = a + f_{i,j,k} $ ,  which requires $u_{i,j,k} \leq b$

  \item  $(u_{i,j,k}-a)^2 + (u_{i,j,k}-b)^2 = f_{i,j,k}^2 h^2$
        \quad  $u_{i,j,k} = \frac{a+b}{2} + \frac{\sqrt{2f_{i,j,k}^2h^2-(b-a)^2}}{2}$ , which requires $u_{i,j,k} \leq c$   (the root number holds because $a + f_{i,j,k}h > b$)

  \item $(u_{i,j,k}-a)^2 + (u_{i,j,k}-b)^2 + (u_{i,j,k}-c)^2 = f_{i,j,k}^2 h^2$
        \quad  $u_{i,j,k} = \frac{a+b+c}{3} + \frac{\sqrt{3f_{i,j,k}^2h^2-(b-a)^2-(c-b)^2-(c-a)^2}}{3}$ 
        (because $\frac{a+b+\sqrt{2f_{i,j,k}^2h^2-(b-a)^2}}{2} > c$, thus $a+b+f_{i,j,k}h > 2c$, thus $a+f_{i,j,k}h > c$)
        
    \vspace{15pt}
    
\end{enumerate}

With the travel time in each specific grid calculated, the method can move on to calculate the overall travel time field. 

Initially, we stipulate that all the grids in this region have an original time field, with zero in the grid where an earthquake occurs and near infinite in other grids. Secondly, as the ideal travel time should be consistent with observation, which represents the first arrival travel time, we need to iteratively update the travel time with a smaller one using travel time in its neighbor grids. To be more specific, the script should have an iteration and simultaneously evaluate if the travel time field achieves stability. Finally, it comes to designing the updating sequence. This method proposes to sweep the whole domain with alternating orderings. In this 3D underground seismic wave propagating case, there are eight orderings in total. 

    \begin{minipage}[t]{0.45\linewidth}
        \begin{enumerate}[label = (\arabic*)]
            \item i = $1:I$,  j = $1:J$,  k = $1:K $
            \item i = $1:I$,  j = $1:J$,  k = $K:1 $
            \item i = $1:I$,  j = $J:1$,  k = $1:K $
            \item i = $1:I$,  j = $J:1$,  k = $K:1 $
        \end{enumerate}
    \end{minipage}
    \hfill
    \begin{minipage}[t]{0.55\linewidth}
        \begin{enumerate}[label = (\arabic*)]
            \setcounter{enumi}{4} 
            \item i = $I:1$,  j = $1:J$,  k = $1:K $
            \item i = $I:1$,  j = $1:J$,  k = $K:1 $
            \item i = $I:1$,  j = $J:1$,  k = $1:K $
            \item i = $I:1$,  j = $J:1$,  k = $K:1 $
        \end{enumerate}
    \end{minipage}
\\
To summarize, we introduce the fast sweeping method in this section, which utilizes the eikonal function to decide the travel time field. Please find the specific script structure in the Algorithm(2.1).
\\
\begin{algorithm}[h]
  \SetAlgoLined
  \KwData{the earthquake's location, the overall velocity model}
  \KwResult{the travel time field of this earthquake}

  initialization: the travel time field should be infinite except grids related to the earthquake
  \While{true}{
    copy and record the old travel time field\;
    \For{i in 8 sweeping algorithm}{
        use each sweeping method to update the new travel time field\;
    }
    \If{the old travel time is the same as the new travel time}{
        break\;
    }
    delete the old travel time field and reserve the new one\;
  }
  \caption{fast sweeping method}
  \label{algo:algorithm1}
\end{algorithm}

\begin{figure}
    \centering
    \includegraphics[width=0.4\columnwidth, height=0.35\linewidth]{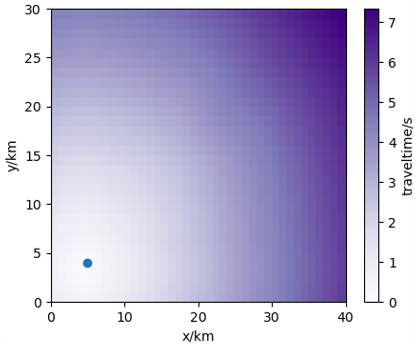}
    \includegraphics[width=0.4\columnwidth, height=0.35\linewidth]{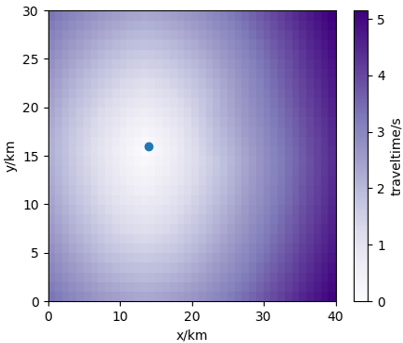}
    \caption{provides examples of travel time field. The blue marker represents the event's location. The value of each grid is calculated through the fast-sweeping method.}
\end{figure}

\section{Inversion Method}
\subsection{Tomography}

After utilizing the fast sweeping method based on the eikonal equation, we can easily obtain the travel time field if we have the exact earthquake location and the 3D velocity model. Thus, the forward process functions now. As we hope to obtain an underground velocity model using observed travel time data, an optimization based on the forward process is necessary. We develop an L-2 loss function of travel time residual and adjust the velocity model to achieve the minimum of this loss function. 
\begin{equation}
L = \sum^{ray path} (u_{obs} - u_{cal})^2
\end{equation}

Calculating the gradient is critical in the optimization. In the tomography part, we try to calculate the gradient in two parts. 
\begin{equation}
    \frac{\partial L}{\partial f} = \frac{\partial L}{\partial u} \cdot \frac{\partial u}{\partial f}
\end{equation}

In this formula, $\frac{\partial L}{\partial u}$ can be calculated through the definition of the loss function. We still use automatic differentiation to make computation easier because the loss function always contains the regularization part, and the original travel time field is always affected by the exact earthquake location. 

As to $\frac{\partial u}{\partial f}$, we calculate this gradient based on the forward equation in the fast sweeping method(equation 2.16).
\begin{equation}
    [(u-a)^+]^2 + [(u-b)^+]^2 + [(u-c)^+]^2 = f^2h^2
\end{equation}

After the differentiation of the equation (3.3), we obtain a differential relationship between the travel time field and the specific velocity model.
\begin{equation}
    (u-a)^+ \cdot d(u-a)^+ + (u-b)^+ \cdot d(u-b)^+ + (u-c)^+ \cdot d(u-c)^+ = h^2f \cdot df
\end{equation}

After rewriting the left hand as, 
\begin{equation}
    [(u-a)^+ + (u-b)^+ + (u-c)^+] \cdot du - (u-a)^+ \cdot da - (u-b)^+ \cdot db - (u-c)^+ \cdot dc
\end{equation}

we can use $ U \cdot \bm{du}$ to represent the left side with $U$ a $ (I*J*K, I*J*K)$  matrix. Here, $U \cdot \bm{du}$ represents matrix multiplication; $\bm{f} \circ \bm{df}$ represents the dot product of two vector elements.
\begin{equation}
    U \cdot \bm{du} = h^2  \bm{f} \circ \bm{df}
\end{equation}

Consequently, we can derive the solution to $\frac{\partial L}{\partial f}$ by solving a matrix equation.
\begin{equation}
    U \cdot \frac{\partial L}{\partial f} = h^2 \bm{f} \circ \frac{\partial L}{\partial u}
\end{equation}

Finally, we obtain the derivative of the loss function $L$ with respect to the velocity model  $f$ by providing $L$'s derivative with respect to the travel time field $u$ and solving the matrix based on the eikonal function. Thus, we can inverse the velocity model by optimizing the loss function with L-BFGS.

\begin{figure}[htbp]
    \centering
    \subfigure{
        \includegraphics[width=0.45\textwidth]{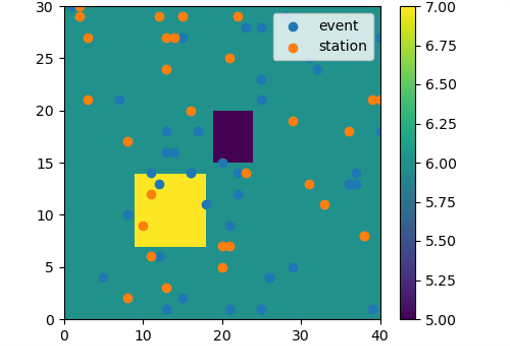}
    }
    \subfigure{
        \includegraphics[width=0.45\textwidth]{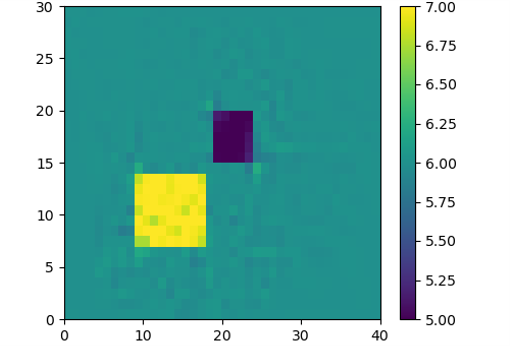}
    }
    \caption{presents a two-dimensional synthetic test using this tomography method. In the first figure, we show the initial velocity model with events in blue and stations in orange on the scatter plot. By combining events and earthquakes' locations with the travel time of each station-event pair, we inverse the velocity model of Figure (b).}
    \label{fig:synthetic_tests}
\end{figure}

\subsection{Relocation}

In the tomography part, we straightforwardly use events' locations from downloaded earthquakes, which does not use an accurate velocity model to locate earthquakes. Thus, relocation is necessary to improve the event catalog and the final velocity model. Regularly, people relocate the seismic events and inverse the velocity model iteratively. 

The main schedule for relocation is to inverse the event's locations using a new inversed velocity model by minimizing the travel time residuals. During the optimization, we need to adjust the event's location and occurance time to reach a minimum of residuals from all the stations. 

Firstly, we continue to rely on the fast sweeping method to build the travel time field. As the travel time from a station to an event should be the same as the counterpart from an event to a station, we build the travel time field based on each station. To be more specific, we set the location of each station to zero and calculate the travel time of each grid in the updated velocity model. In the inversion part, we define the loss function and the gradient, which are essential components that enable us to achieve inversion.
\begin{equation}
    L = \frac{1}{2}\sum^{ray path} (t_{obs}-t_{cal})^2
\end{equation}

Because the variables are the locations of an event and its happening time, we are interested in this gradient $(\frac{\partial L}{\partial x}, \frac{\partial L}{\partial y}, \frac{\partial L}{\partial z}, \frac{\partial L}{\partial t})$.

If we obtain the travel time through one constant velocity value and the distance between the station and the event, 
\begin{equation}
    t_{cal} = \frac{\sqrt{(x_{sta}-x_{eve})^2+(y_{sta}-y_{eve})^2+(z_{sta}-z_{eve})^2}}{v}
\end{equation}

then we can obtain the gradient of $L$ immediately. (We use $x_1$, $x_2$, and $x_3$ to represent $x$, $y$, and $z$ for simplification.)
\begin{equation}
    \frac{\partial L}{\partial x_{i}} = \sum^{ray path} \frac{\partial L}{\partial t} \cdot \frac{\partial t}{\partial x_i} = \sum^{ray path} (t_{cal}-t_{obs}) \frac{x_{i,event}-x_{i,station}}{v\sqrt{\Delta x_{1}^2 + \Delta x_{2}^2 + \Delta x_{3}^2}},
    \quad i = 1,2,3
\end{equation}

\begin{equation}
    \frac{\partial L}{\partial t} = \sum^{ray path} (t_{cal}-t_{obs})
\end{equation}

In order to apply this function to a more realistic circumstance, we need to use a more sophisticated velocity model. The calculation of the travel time field and the loss function still uses the fast sweeping method. Simultaneously, we need to provide a gradient matrix for future optimization. In the initialization part, we calculate the gradient of the travel time field in different dimensions using a function in NumPy. 
Thus, we obtain 
$\frac{\partial t_{i,j,k}}{\partial x_i}$, $\frac{\partial t_{i,j,k}}{\partial x_j}$, $\frac{\partial t_{i,j,k}}{\partial x_k}$ of each grid in the whole research region, which could a straightforward calculation of the loss function's derivatives with regard to locations.

Regularly, the earthquake's location is not on the grid point, so interpreting is necessary to obtain the travel time's gradient in an exact earthquake's location. For simplification, we can use another expression to represent the derivative of the travel time on each grid with regard to different directions.
\begin{equation}
g_1(i,j,k) = \frac{\partial t_{i,j,k}}{\partial x}, \quad g_2(i,j,k) = \frac{\partial t_{i,j,k}}{\partial y}, \quad g_3(i,j,k) = \frac{\partial t_{i,j,k}}{\partial z}
\end{equation}

If we assume the event's location $(x,y,z)$ satisfies,
\begin{align}
    x_0 + i*h<\bm{x}<x_0 + (i+1)*h\\
    y_0 + j*h<\bm{y}<y_0 + (j+1)*h\\
    z_0 + k*h<\bm{z}<z_0 + (k+1)*h
\end{align}

we can obtain the derivative in the event's location with regard to $x$, namely $g_1(x,y,z)$, using the linear interpreting method to process the ambient known gradient values $g_1(i,j,k)$, $g_1(i,j,k+1)$, $g_1(i,j+1,k)$, $g_1(i,j+1,k+1)$, $g_1(i+1,j,k)$, $g_1(i+1,j,k+1)$, $g_1(i+1,j+1,k)$, $g_1(i+1,j+1,k+1)$. The loss function's gradients with regard to other directions, $g_2(x,y,z)$ and $g_3(x,y,z)$, can also be obtained using the same algorithm.



Then, we can use the loss function and the gradient for optimization. During the optimization part, direct optimization of the loss function works if the picks we use are accurate and actually from the event we research. However, in most cases, the selections may deviate or even be unreasonable. Therefore, we use RANSAC to automatically utilize reasonable picks and then achieve a more accurate inversion.

\begin{figure}[htbp]
    \centering
    \subfigure{
        \includegraphics[width=0.48\textwidth]{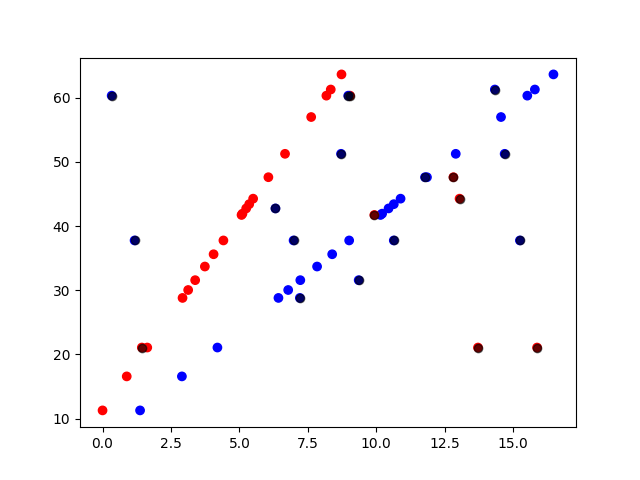}
    }
    \subfigure{
        \includegraphics[width=0.48\textwidth]{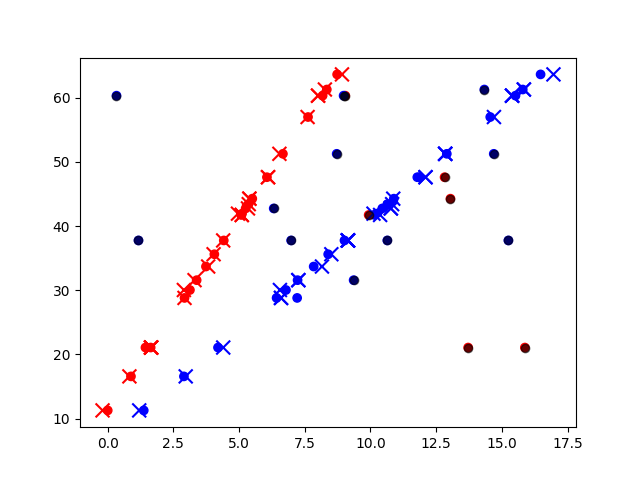}
    }
    \caption{presents a synthetic test of relocation. The first subfigure displays synthetic picks generated from one event and several stations with some noise picks. Red circles represent synthetic P wave picks, and blue circles represent synthetic S wave picks. Circles in dark colors are noise data. The second subfigure displays the travel time to each station based on the inversed event's location from picks in Figure(a). We use crosses to represent picks from the inversed location.}
    \label{fig:synthetic_tests}
\end{figure}

\section{Results}

In this chapter, we will present the realistic applications of this method. We use this novel method to obtain the velocity model as well as the Vp/Vs ratio in the San Francisco Bay Area. 

\subsection{Data Preparation}
In the realistic inversion, we choose a region in the San Francisco Bay Area. The region of this dataset spans from $120^{\circ} \text{W}$ to $124^{\circ} \text{W}$ in longitude, and from $36^{\circ} \text{N}$ to $38.6^{\circ} \text{N}$ in latitude.  

\begin{figure}[htbp]
    \centering
    \includegraphics[width=0.6
    \textwidth]{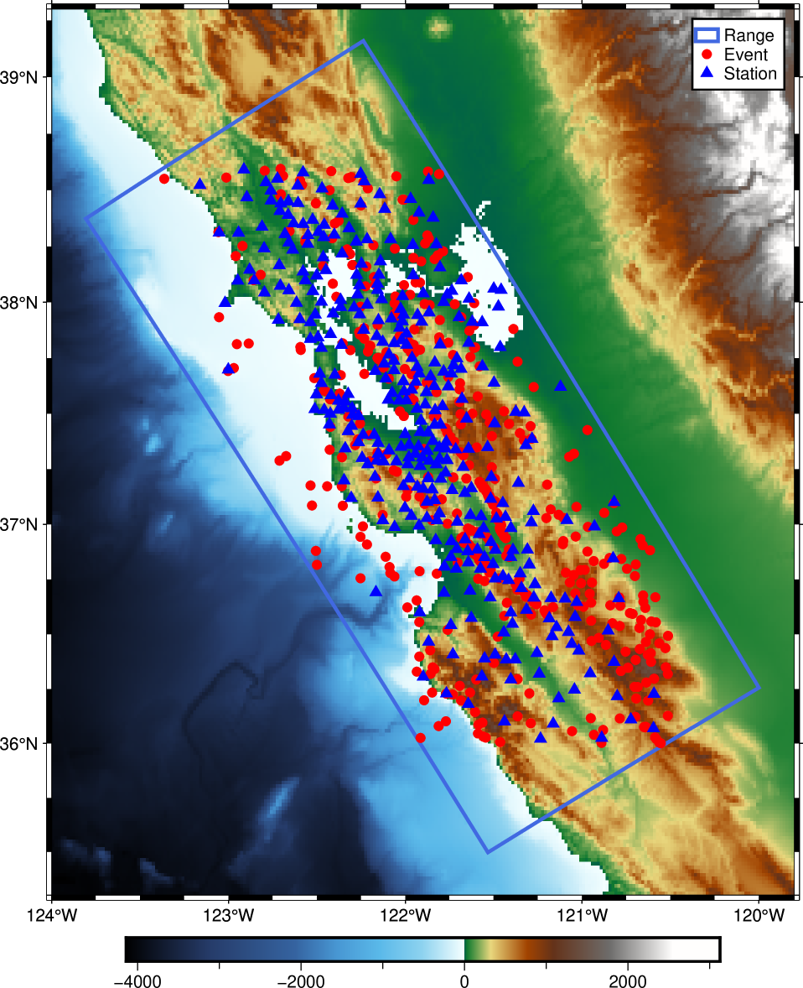}
    \caption{presents the research range in the topography map. The royal blue rectangle describes the research range; the blue triangle represents the stations used in this project; the red circle represents the events used. }
    \label{4.1}
\end{figure}

We do not use any active sources here, and all the datasets we use come from natural earthquakes. In the time range, we use events from 2009.01.01 to 2023.01.01. This dataset has 643 stations and 5565 events in total. From these events, we obtain 350,469 P wave picks and 199,479 S wave picks utilizing PhaseNet(\citet{zhu2019phasenet}). We use distances of $km$ to research this region and then convert the coordinates with a degree of $32^{\circ}$. Finally, we obtain a research region of $164km \times 380km \times 30km$, with a minimum resolution of $2km$. Because of the limited computation resources, we use half of the total stations and almost one-tenth of the total events. 

Before inversion, we plot the residuals between calculated travel time and observed travel time of each station and try to have a direct understanding of shallow underground layers. We calculate seismic travel time through locations of stations and events with the GIL7 1D velocity model(\citet{stidham1999three}). Using this formula,
\begin{equation}
    r_i = \frac{num_{\text{obs} > \text{cal}}-num_{\text{obs} < \text{cal}}}{num_{\text{obs} > \text{cal}} + num_{\text{obs} < \text{cal}}}
\end{equation}

we then obtain the average level of residual in this region. In this equation, $r_i$ represents the residual level of station i; $num_{\text{obs} > \text{cal}}$ represents the number of picks in station i with longer observed travel time; $num_{\text{obs} < \text{cal}}$ represents the number of picks in station i with longer calculated travel time. 

\vspace{5pt}
\begin{figure}[htbp]
    \centering
    \includegraphics[width=1
    \textwidth]{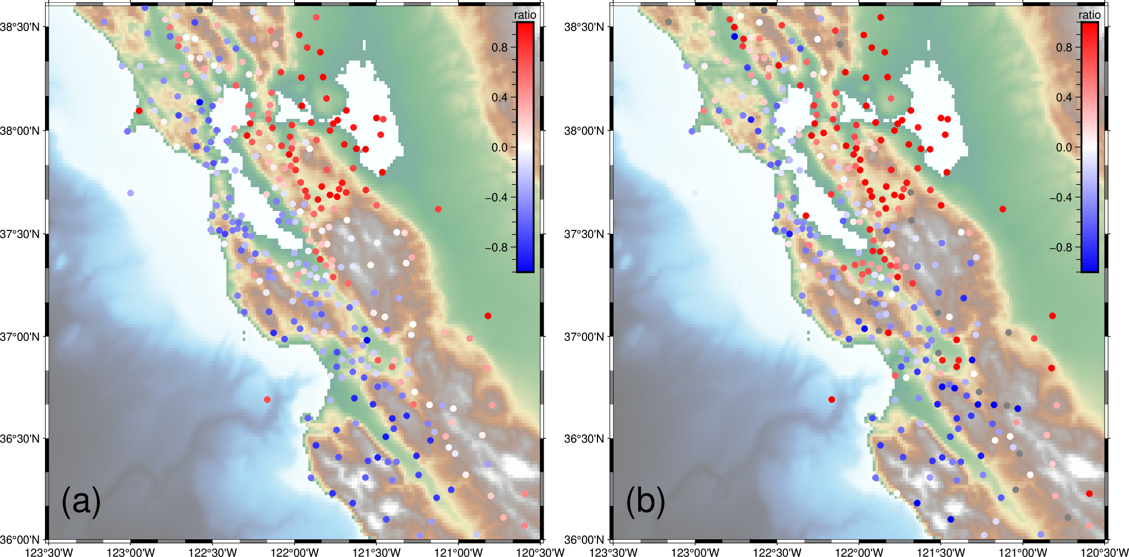}
    \caption{presents the residuals before inversion. Each circle represents a station with color equaling the residual value $r_i$ of each station. If the calculated travel time in one station is often longer, then $r_i$ will be a positive value.}
    \label{residual}
\end{figure}

We plot these residual levels using the red-blue color bar. In Figure \ref{residual}, the redder the point, the lower the velocity in that region, as the observed travel time is longer than the calculated travel time. Correspondingly, negative $r_i$ represents longer observed travel time. The bluer the point, the higher the velocity in that region, as the observed travel time is shorter than the calculated travel time.

\subsection{Coverage Tests}

In this section, we apply some ray path coverage tests and checkerboard tests to evaluate the dataset's coverage. 

We plot the ray path of each pick from the event to the station. In our research region, we track the number of ray paths passing through each grid. We then create a heat map to plot the logarithm of these counts. From the data coverage, we probably have better resolution in the middle and shallow parts of our research region. 

\begin{figure}
    \centering
    \subfigure{
        \includegraphics[width=1\textwidth]{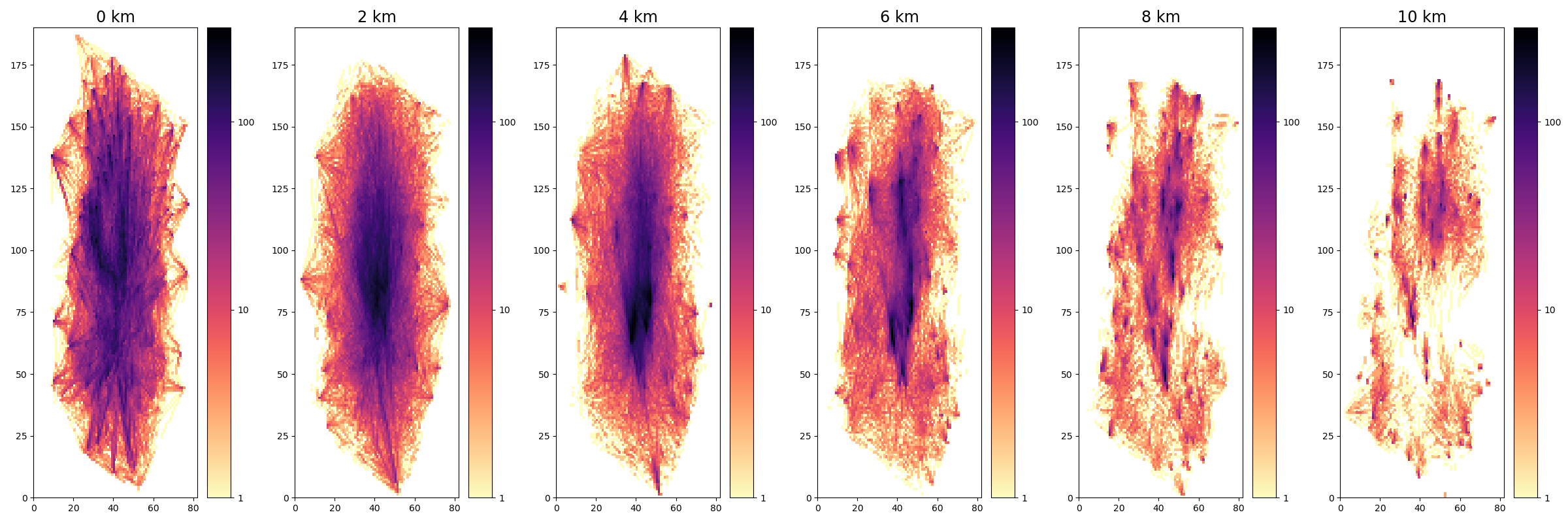}
    }
    \subfigure{
        \includegraphics[width=1\textwidth]{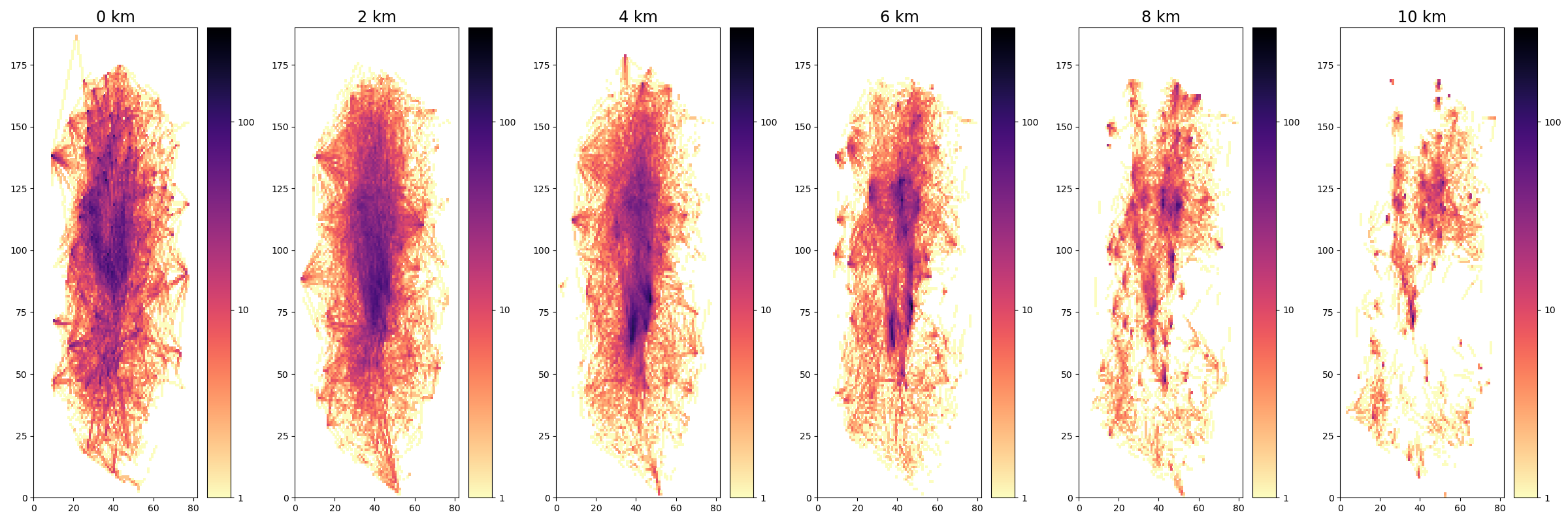}
    }
    \caption{shows the ray path coverage.(a) shows the ray path coverage of the P wave. The color in the map means the logarithm of the passing P wave counts. Figure \ref{cov}(b) shows the ray path coverage of the S wave, which has the same meaning as Figure \ref{cov}(a).}
    \label{cov}
\end{figure}



Then, we use the checkerboard tests in this region to further ensure the coverage of ray path. In the synthetic checkerboard test, we set the length of the velocity block to be ten grids. Using events and stations the same as the inversion process, we generate P wave and S wave picks from designed velocity model. We use travel time residuals in the synthetic model and appropriate regularization to inverse the velocity model, thus having a better understanding of inversion resolution. 

\begin{figure}
    \centering
    \subfigure{
        \includegraphics[width=1\textwidth]{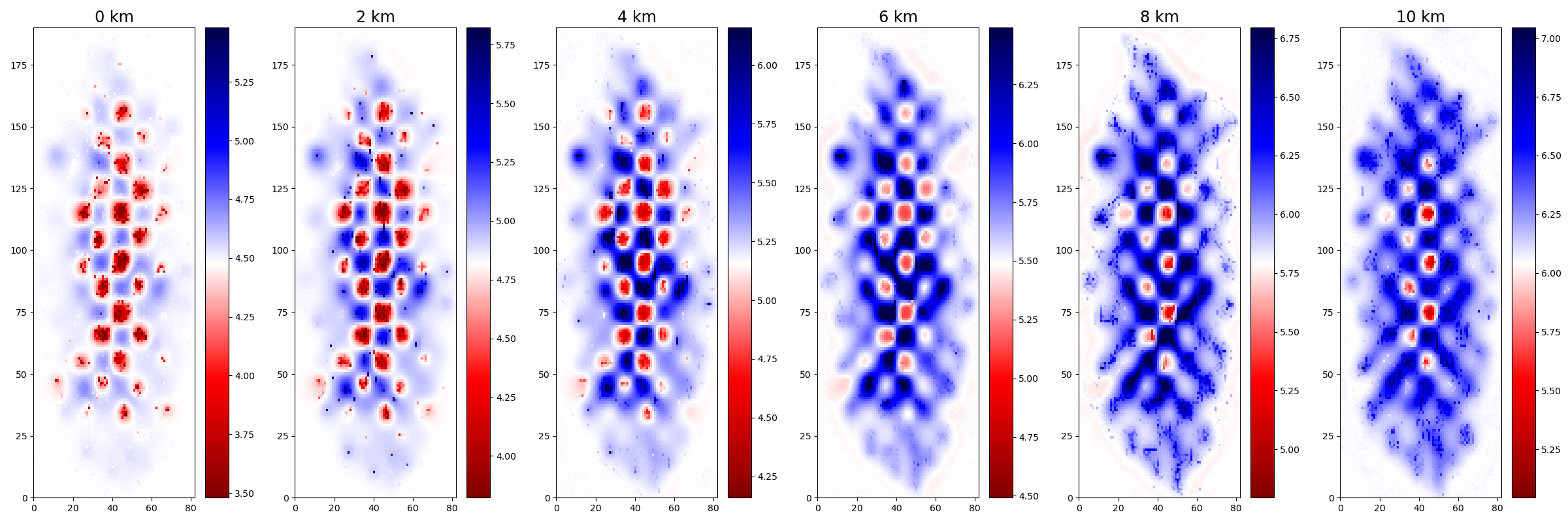}
    }
    \subfigure{
        \includegraphics[width=1\textwidth]{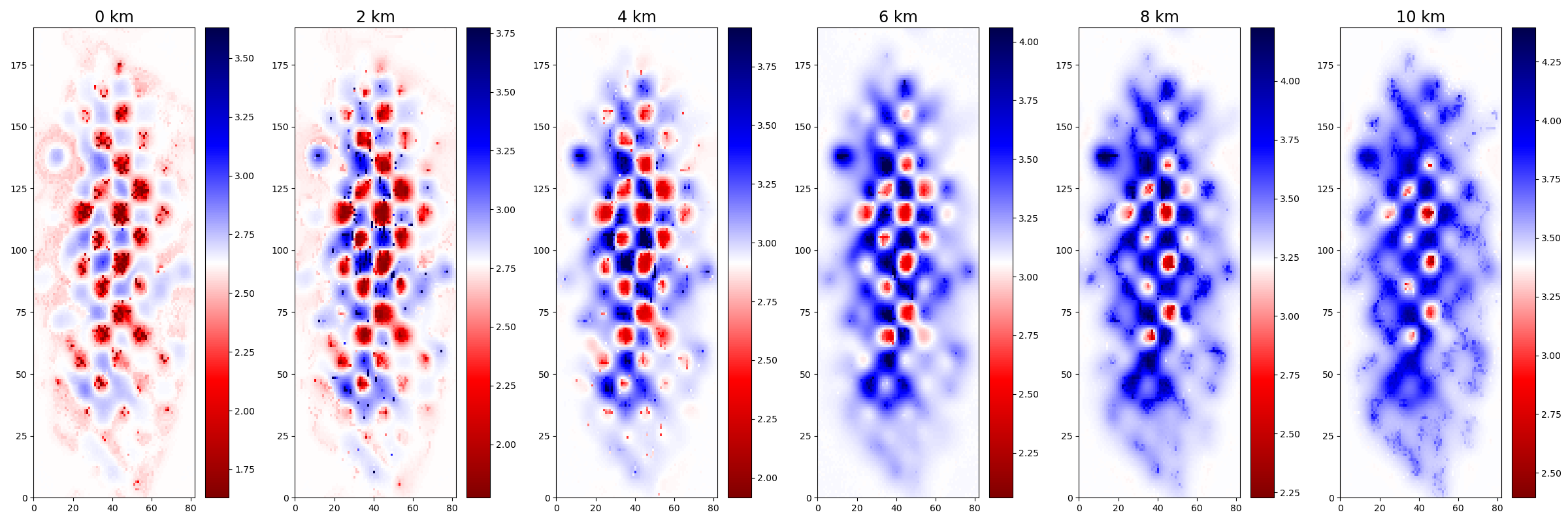}
    }
    \caption{presents the checkerboard test results. (a) presents the checkerboard test results of the P wave; Figure \ref{check}(b) presents the checkerboard test results of the S wave. Each unit figure represents a layer with the same depth in the total velocity model. }
    \label{check}
\end{figure}

From Figure \ref{check}, The P wave velocity model has a higher resolution than the S wave velocity model.

\subsection{Inversion Results}

The checkerboard test results show that the depth resolution is not satisfying. This issue may stem from limitations in both inversion resolution and computational resources. Another possible reason is the constrained observation system, especially the station locations. As to the realistic results, the dataset with this method may not obtain a very detailed depth resolution. 

In this section, we present figures for the inversion results. As the inversion result of the P wave and S wave are similar to each other, we display the P wave velocity result and the Vp/Vs ratio. This inversion starts from the GIL7 velocity model in the paper(\citet{stidham1999three}).

\vspace{5pt}
\begin{figure}[htbp]
    \centering
    \includegraphics[width=1
    \textwidth]{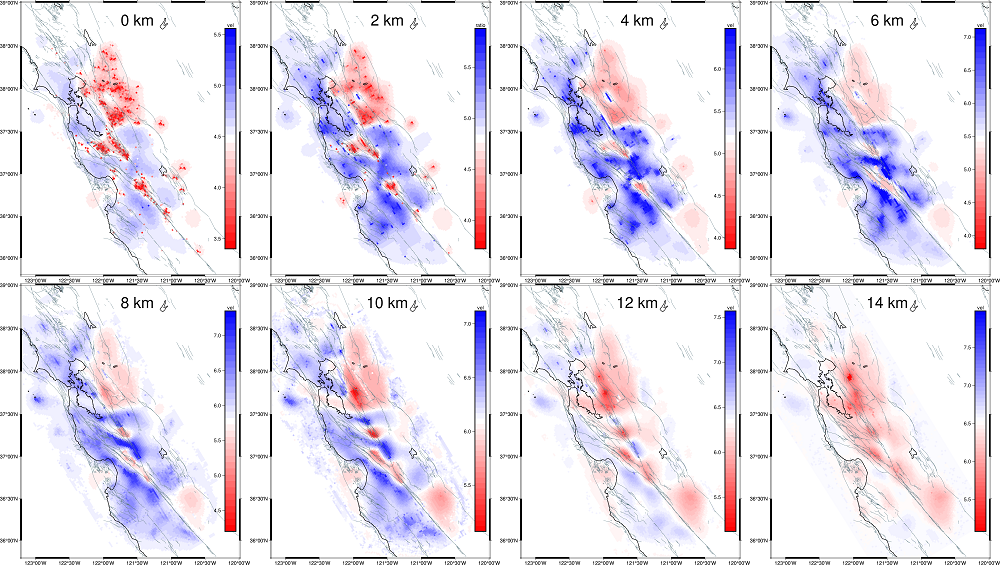}
    \caption{presents the results of P wave velocity model after inversion, with each unit figure representing a layer with the same depth. Each figure also consists of fault zone architecture.}
    \label{velp}
\end{figure}

\begin{figure}
    \centering
    \subfigure{
        \includegraphics[width=0.35\textwidth]{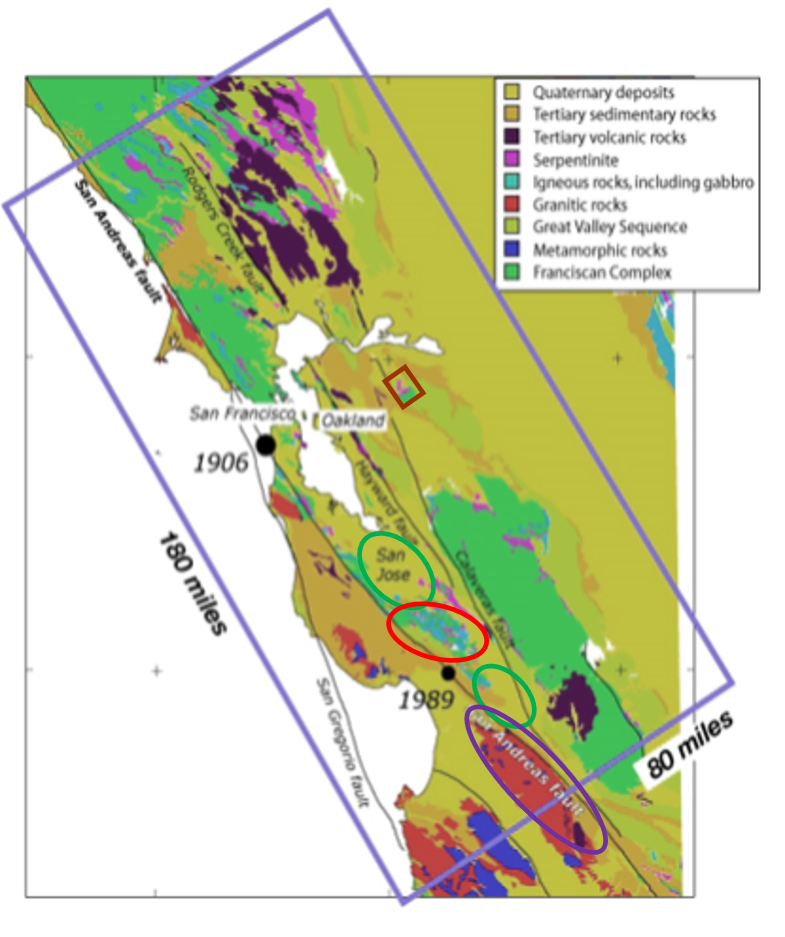}
    }
    \subfigure{
        \includegraphics[width=0.6\textwidth]{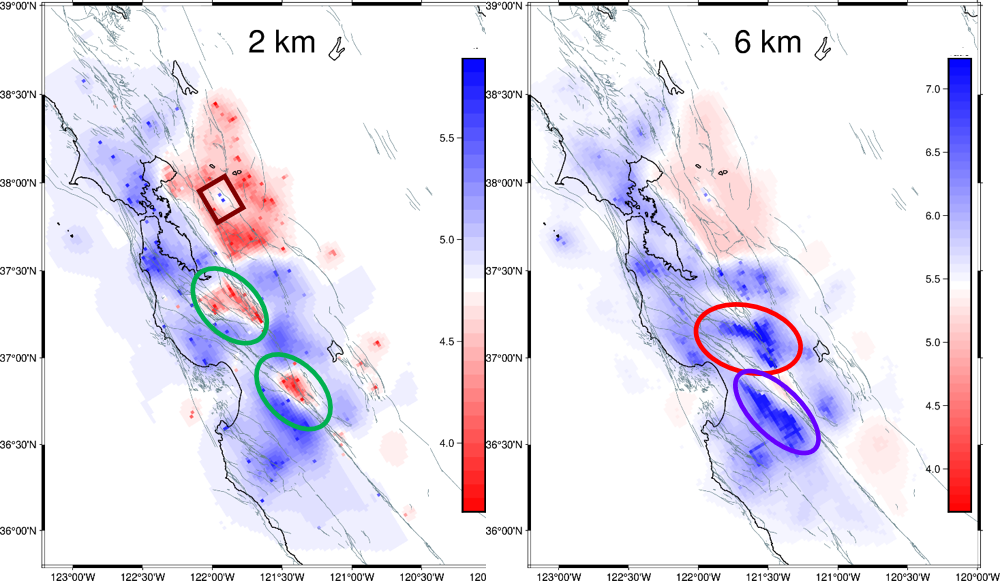}
    }
    \caption{includes the geological map from USGS(a) and two velocity unit figures in shallow layers(b). Different markers, like circles and triangles, in the geological map remain consistent with the markers at the exact locations in the velocity figures. Velocity anomalies thus can be explained by geological evidence. }
    \label{usgs}
\end{figure}

We compare our results with P wave velocity model from other papers and find that the velocity model provided here exhibits more details than previous results. To further examine the correctness of the results in this paper, a comparison with geological maps is necessary. In Figure \ref{usgs}, we use the geological map of the San Francisco Bay Area from USGS(\citet{graymer2006geologic}) to explain the low-velocity anomaly and high-velocity anomaly in the inversion result.

By comparing the velocity model with the geological map, we can efficiently conduct an analysis. Generally, the velocity in the eastern region is slower than in the western region. On a more detailed level, petrology can explain high and low-velocity anomalies. A high-velocity anomaly within a low-velocity region is because serpentinite, a metamorphic rock, is present in a large area of deposits and sedimentary rocks. The presence of Quaternary deposits and sedimentary rocks explains the low-velocity regions near the San Andreas Fault and Hayward Fault. The high-velocity anomalies in these regions are due to igneous rocks in the geological map.

\begin{figure}[htbp]
    \centering
    \subfigure{
        \includegraphics[width=0.31\textwidth]{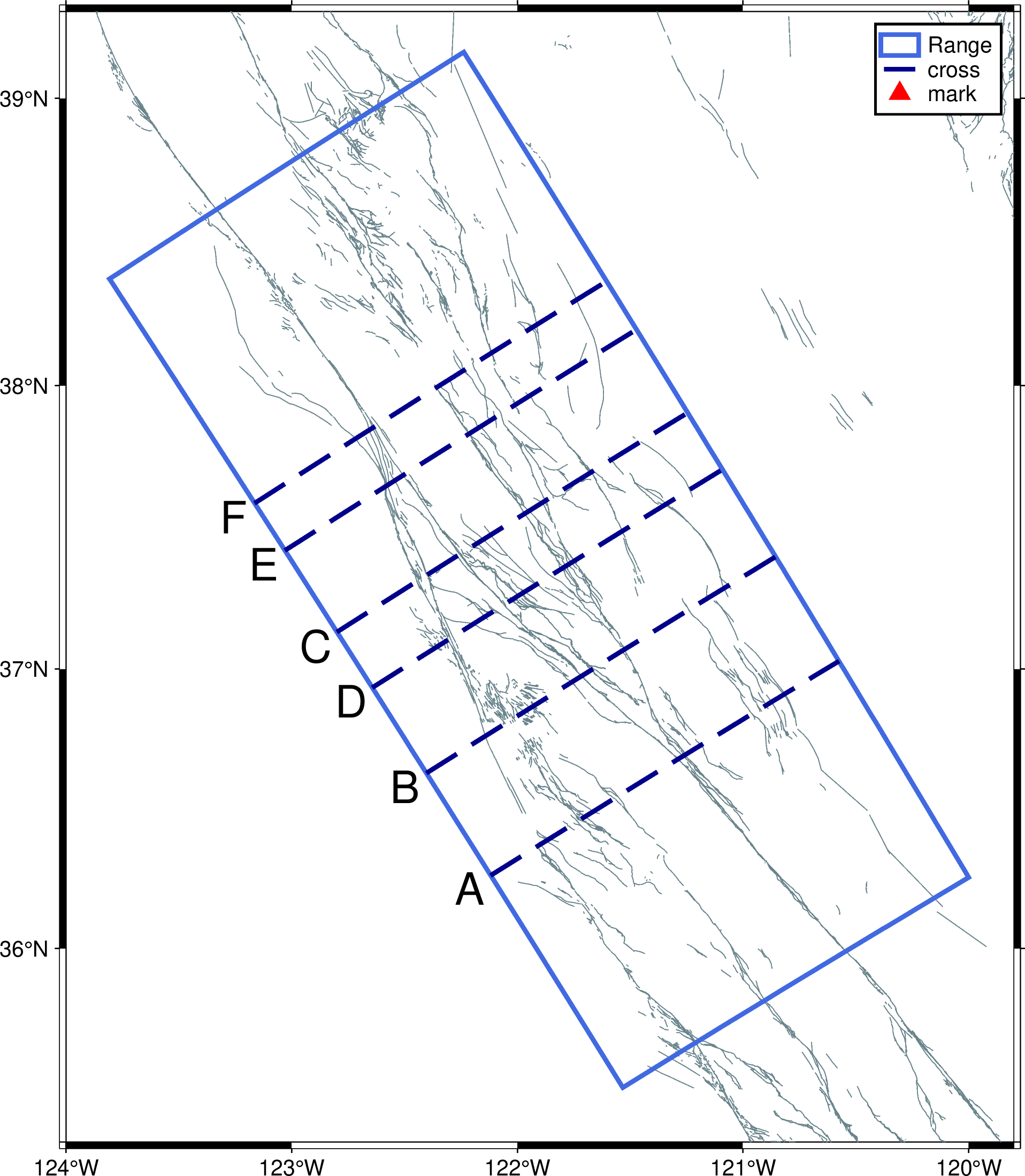}
    }
    \subfigure{
        \includegraphics[width=0.31\textwidth]{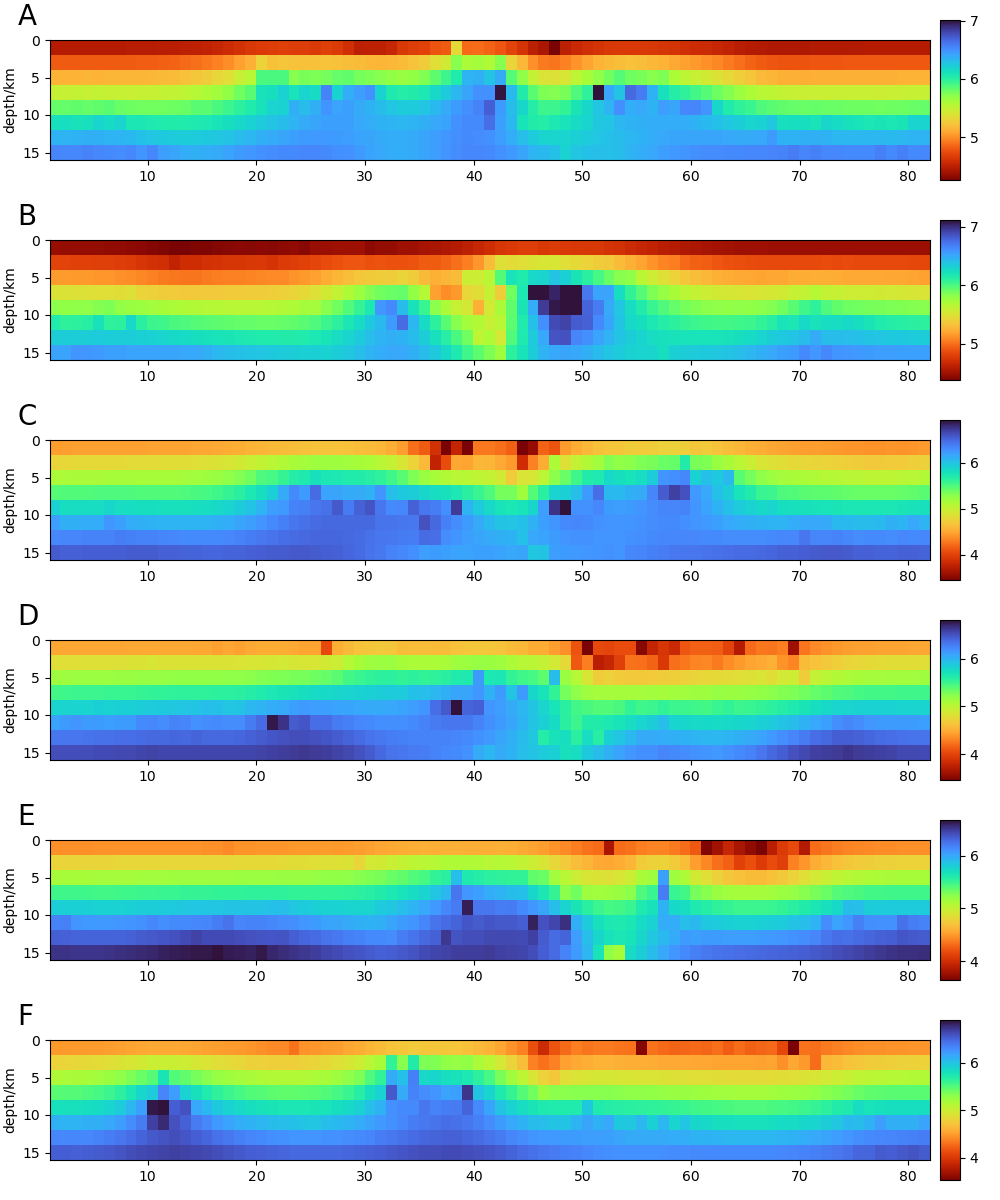}
    }
    \subfigure{
        \includegraphics[width=0.31
        \textwidth]{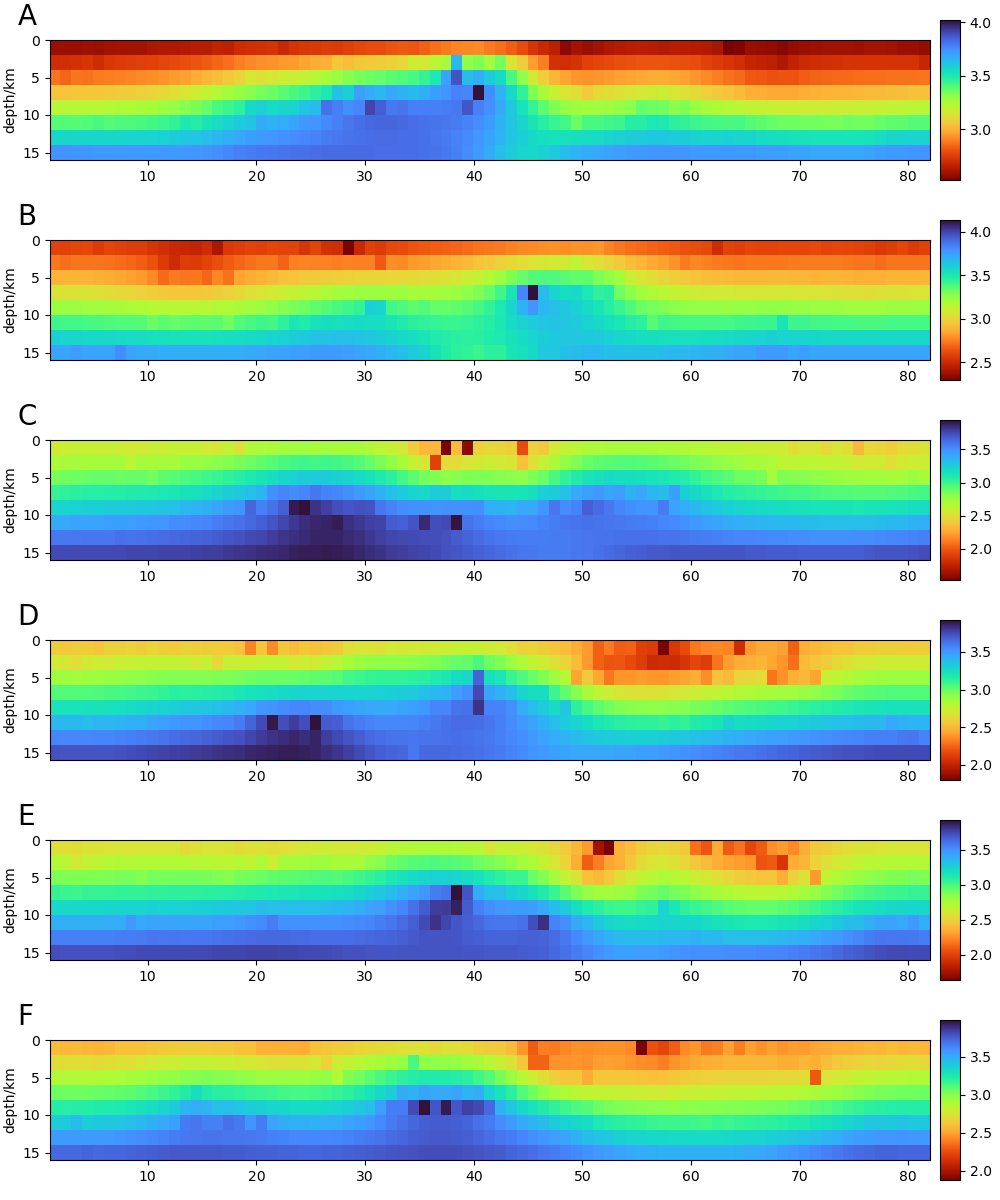}
    }
    \caption{plots the cross-sections vertical to the fault direction. Figure (a) exhibits the horizontal locations of these cross-sections on the research range with fault zone architecture. Figure (b) plots the P wave's cross-sections; Figure (c) plots the S wave's cross-sections; both of the unit figures correspond to figure(a) with labels(A-F).}
    \label{crossy}
\end{figure} 

\begin{figure}[h]
  \begin{minipage}[b]{0.52\textwidth}
    \centering
     \subfigure{\label{Genelecs:Genelec 8020 AP}\includegraphics[width=\textwidth]{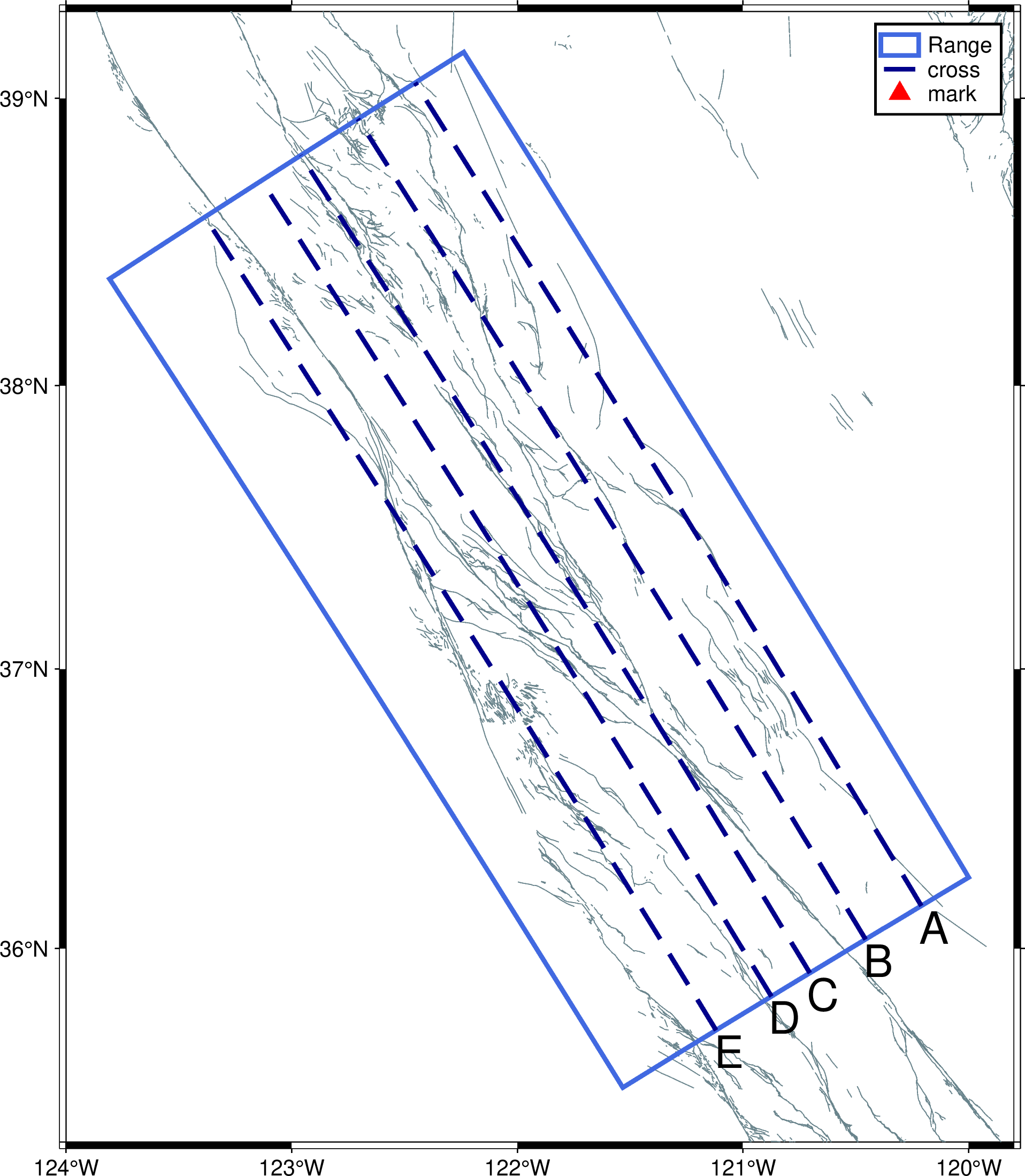}}
    
  \end{minipage}%
  \hfill
  \begin{minipage}[b]{0.45\textwidth}
    \centering
     \subfigure{\label{Genelecs:Genelec 8020 AP}\includegraphics[width=\linewidth]{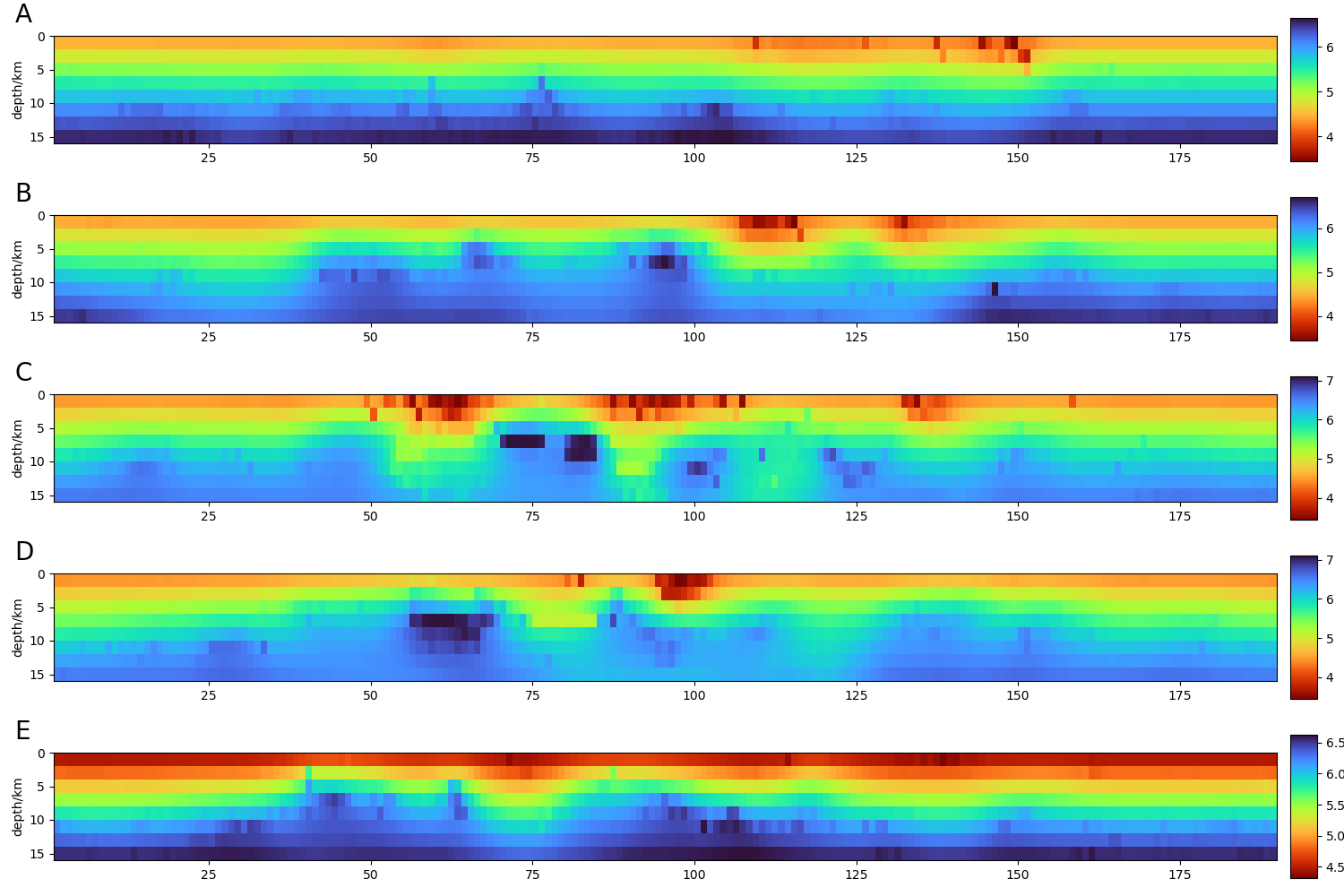}}
     \subfigure{\label{1}\includegraphics[width=\linewidth]{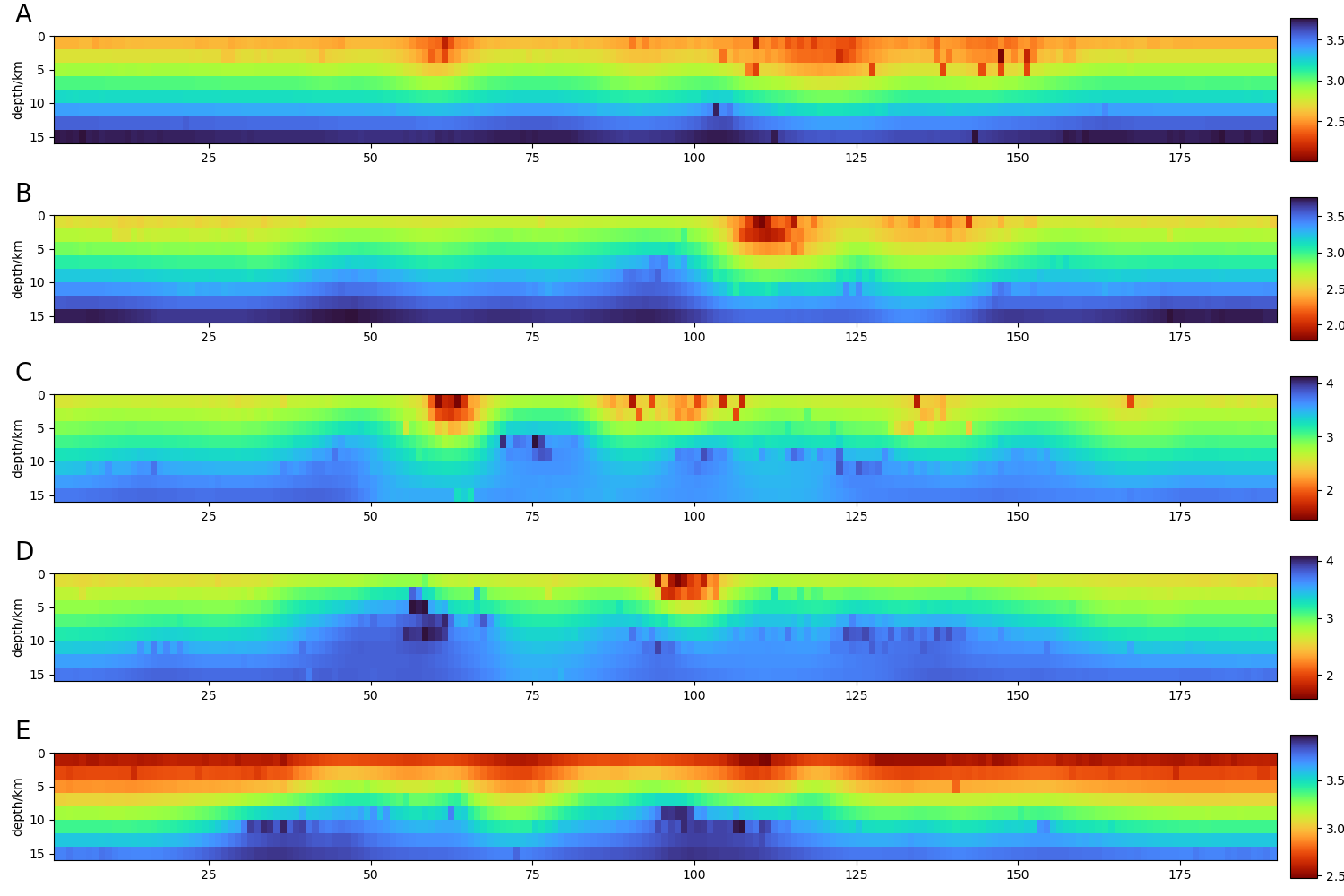}}
 
  \end{minipage}
  \caption{is similar to Figure \ref{crossy} but plots the cross-sections parallel to the fault direction. (a) exhibits cross-sections' locations; (b) and (c) plot cross-sections of respective waves, with labels(A-E).}
  \label{crossx}
\end{figure}

In figure \ref{crossy} and figure \ref{crossx}, we plot the cross-section results to find the relationship between velocity and depth. In these two cross sections, both P wave and S wave models exhibit velocity contrast in regions near fault zones. In the cross-sections vertical to the fault zones, the P wave and S wave velocity models are very similar to each other. In the cross-sections parallel to the fault zones, the P wave and S wave velocity models exhibit some apparent differences when the horizontal location is close to fault zones. Therefore, it is possible to probe that the Vp/Vs velocity ratio in fault regions deviates from the standard value. 

We obtain the Vp/Vs ratio through joint inversion. In this process, the variables contain the P wave velocity model and Vp/Vs ratio, which are optimized simultaneously. The loss function is still composed of the travel time residual and the regularization part. We use equation \ref{joint} to describe the loss function composed of travel time residual. 
\begin{equation}
    L_res = \sum^{P} (t_{syn,P} - t_{obs,P})^2 + \sum^{P\& S} (t_{syn,d} - t_{obs,d})^2
    \label{joint}
\end{equation}

Firstly, it should be composed of P wave travel time residual, which helps obtain a reliable P wave velocity model. Besides, we use the differential travel time between S wave travel time and P wave travel time to inverse the Vp/Vs ratio. Traditionally, people assume that the P wave and S wave have the same travel path and use the differential time in this path to inverse related Vp/Vs ratio. Unfortunately, this assumption introduces many errors. In our joint inversion, we calculate the P wave and S wave travel time field independently and then subtract P wave travel time from S wave travel time to obtain the differential travel time. As P wave velocity and Vp/Vs ratio can generate the S wave velocity model, we can reach the minimum differential travel time residual by adjusting the velocity ratio.

Apart from the travel time residual, the loss function is also composed of the regularization part. The inversion process adds smoothing to P wave velocity model, S wave velocity model, and the Vp/Vs ratio. Because the P wave velocity model and the S wave velocity model look similar, the velocity ratio should be a constant value in most areas with abrupt values in several small regions. Through carefully designed loss function and variable change range, we obtain the velocity ratio using joint inversion of P wave and Vp/Vs ratio simultaneously. 

\vspace{5pt}
\begin{figure}[h]
    \centering
    \includegraphics[width=1
    \textwidth]{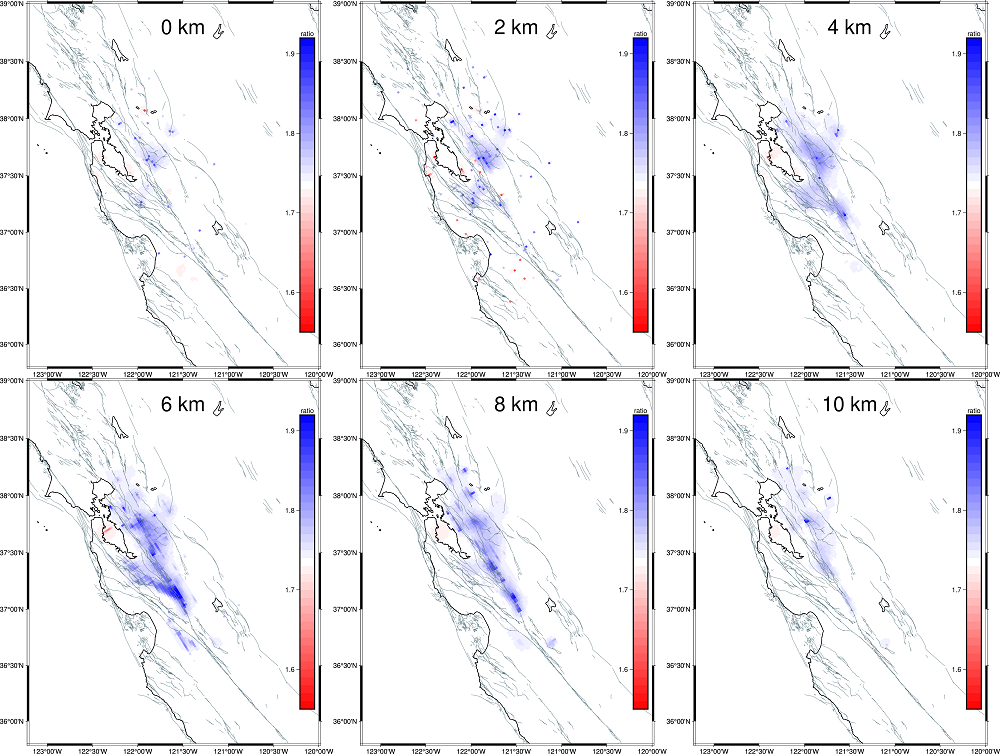}
    \caption{presents the P wave to S wave velocity model, with each unit figure representing a layer with the same depth. These six unit figures have the same color bar and color range. The max depth is 10 km because the Vp/Vs value becomes constant in deeper layers.}
    \label{pvs}
\end{figure}

The respective inversion processes of the P wave and S wave are different, which means calculating the velocity ratio through direct division could introduce or even amplify errors from both models. Thus, Vp/Vs results from joint inversion are much more reliable than those obtained by simply dividing the P wave and S wave velocity models. After we obtain the Vp/Vs ratio result from joint inversion, we compare it with the Vp/Vs model derived from direct division. We found that both models showed similar patterns in high and low velocity ratio anomalies, while results from the joint inversion look reasonable. Besides, the Vp/Vs ratio result tends to remain constant in deep layers, so we only present the shallower layers in Figure \ref{pvs}. 

\section{Conclusions}

In data preparation, this project utilizes a deep learning method to obtain seismic arrival picks automatically, thus providing sufficient travel time data with both P wave and S wave for inversion. 

This project also develops a novel inversion method. We employ the fast sweeping method to solve the eikonal equation, thereby obtaining the travel time field. In the inversion process, we use automatic differentiation and matrix solving to calculate gradients. Notably, we calculate the Vp/Vs ratio results without any approximations. This method can be readily applied to address tomography problems in various regions.

As to the inversion results, the highly resolved P wave velocity model in this paper is much more similar to the geological map than previous results because of the abundant ray path coverage and this new method. Besides, we also obtain the S wave velocity model and Vp/Vs ratio results of the San Francisco Bay Area, enabling a complete understanding of seismic properties. Our results also display distinct velocity contrast within Fault Zones, helping us learn more about these crucial geological areas.

\end{document}